# Riemannian Geometry for the classification of brain states with intracortical brain-computer interfaces

Arnau Marin-Llobet[1,2], Arnau Manasanch[1,3], Sergio Sánchez-Manso[4], Lluc Tresserras[1,5], Xinhe Zhang[2], Yining Hua[6,7], Hao Zhao[2], Melody Torao-Angosto[1], Maria V Sanchez-Vives[1,8], Leonardo Dalla Porta[1]

[1] Institut d'Investigacions Biomediques August Pi i Sunyer (IDIBAPS), Barcelona, Spain
[2] John A. Paulson School of Engineering and Applied Sciences, Harvard University, Boston, MA, USA
[3] Faculty of Medicine and Health Sciences, University of Barcelona, Barcelona, Spain
[4] Whiting School of Engineering, Johns Hopkins University, Baltimore, MD, USA
[5] School of Mathematics and Statistics, Polytechnic University of Catalonia, Barcelona, Spain
[6] Harvard T.H. Chan School of Public Health, Boston, MA, USA
[7] Beth Israel Deaconess Medical Center, Harvard Medical School, Boston, MA, USA
[8] Catalan Institution for Research and Advanced Studies (ICREA), Barcelona, Spain
amarinllobet@seas.harvard.edu , msanche3@recerca.clinic.cat, dallaporta@recerca.clinic.cat

## Abstract

This study investigates the application of Riemannian geometry-based methods for brain decoding using invasive electrophysiological recordings. Although previously employed in non-invasive, the utility of Riemannian geometry for invasive datasets, which are typically smaller and scarcer, remains less explored. Here, we propose a Minimum Distance to Mean (MDM) classifier using a Riemannian geometry approach based on covariance matrices extracted from intracortical Local Field Potential (LFP) recordings across various regions during different brain state dynamics. For benchmarking, we evaluated the performance of our approach against Convolutional Neural Networks (CNNs) and Euclidean MDM classifiers. Our results indicate that the Riemannian geometry-based classification not only achieves a superior mean F1 macro-averaged score across different channel configurations but also requires up to two orders of magnitude less computational training time. Additionally, the geometric framework reveals distinct spatial contributions of brain regions across varying brain states, suggesting a state-dependent organization that traditional time series-based methods often fail to capture. Our findings align with previous studies supporting the efficacy of geometry-based methods and extending their application to invasive brain recordings, highlighting their potential for broader clinical use, such as brain computer interface applications.

## Introduction

Brain-computer interfaces (BCIs) establish a direct communication link between brain signals and external devices. Also known as brain-machine interfaces (BMIs), these systems function as neural decoders, translating neuronal activity into commands for assistive technologies (1,2). BCIs have revolutionized healthcare, with significant impacts on motor control, communication, and neurorehabilitation (3). Notable applications include robotic arm control for paralyzed individuals, speech prostheses for those with severe disabilities, or neurofeedback therapies and virtual agents (4–8). A typical BMI system consists of two primary components: (i) neural signal acquisition and (ii) decoding, often followed by an actuator or feedback mechanism.

The first component, neural signal acquisition, can be broadly categorized as non-invasive or invasive (9). Non-invasive techniques, such as electroencephalography (EEG) (10), functional magnetic resonance imaging (fMRI) (11), and functional near-infrared spectroscopy (fNIRS) (12), are more accessible and less intrusive. On the other hand, invasive methods, such as intracortical local field potentials (LFPs) and high-density single-unit recordings, provide direct access to neural activity with both high spatial and temporal resolution (2,14,15). These invasive signals are essential for capturing fine-grained neural dynamics underlying motor control and complex cognitive processes (16,17). However, their acquisition poses significant challenges, requiring costly and highly specialized surgical procedures, long-term data collection, and ongoing patient care (18,19). These constraints severely limit the availability of training data for deep learning models, restricting the development and validation of advanced neural decoding algorithms.

The second component, the decoder, typically utilizes statistical, signal processing, and/or machine learning algorithms to interpret brain signals, regardless of the acquisition method. Decoding is crucial because neural signals, particularly those from invasive recordings, are highly non-stationary and vary based on brain region, task, and individual state. The standard approach in clinical applications is template matching, which aligns neural data with subject-specific templates (20,21). While effective, this method is constrained by its reliance on predefined templates, limiting its generalizability across individuals. It is also sensitive to noise and artifacts, oversimplifies the nonlinear structure of neural data, and requires frequent recalibration to adapt to new subjects or changing conditions. Recent advancements have introduced deep learning methods, particularly Convolutional Neural Networks (CNNs), which can automatically learn spatial and temporal features from neural data (22–25). However, CNNs and other architectures such as Recurrent Neural Networks (RNNs) face significant challenges, including the need for large datasets, long training times, and GPU support. These factors make them impractical for clinical applications requiring real-time decoding and adaptability. Furthermore, deep learning models, which rely on traditional data processing paradigms, struggle with issues such as feature redundancy, gradient instability, and inefficiency when confronted with the complex, nonlinear nature of neural signals.

To overcome these limitations, geometric-based techniques have gained traction by capturing the intrinsic structure of neural data, often represented as covariance matrices. Unlike conventional methods, geometric approaches exploit the manifold structure of covariance matrices, enhancing generalizability and stability in clinical applications (26). However, since these matrices reside on a non-Euclidean space, standard machine learning algorithms, such as the Minimum Distance to Mean (MDM) classifier, cannot be directly applied. One naive approach is to ignore the manifold's curvature and apply Euclidean methods, but this often results in poor accuracy and computational inefficiencies (27,28).

To better handle the non-stationarity of neural signals, Riemannian geometry-based classification (RGBC) has emerged as a promising alternative, particularly for non-invasive motor activity recordings (29,30). By leveraging geodesic distances between covariance matrices, RGBC provides a more flexible and robust framework for analyzing neural dynamics across individuals and conditions. The geodesic path $\gamma(t)$ represents the shortest trajectory between two points on the manifold $\mathcal{M}$, preserving the intrinsic structure of the data (31). At each point $p$ on $\mathcal{M}$, an inner product is defined in the associated tangent space $\mathcal{T}_p\mathcal{M}$, which is locally Euclidean. This allows computations to be performed efficiently while respecting the manifold's structure (27,28).

Despite the promising results of RGBC, it has been almost exclusively validated on non-invasive signals such as EEG, fNIRS, or fMRI (29,32,33). Its application to invasive brain data remains largely unexplored. This study addresses this gap by investigating the effectiveness of RGBC in classifying local field potentials (LFPs) recorded across multiple brain regions during anesthesia, a challenge that has remained largely unexplored despite its clinical relevance. LFP classification presents a particular challenge, as its data is often highly dynamic and variable across regions. With geometric techniques, we aim to investigate whether RGBC can capture the underlying spatiotemporal patterns of neural activity, effectively differentiating between distinct brain states and accounting for the variability across regions, ultimately providing more robust and accurate decoding of these complex signals. Specifically, this study is guided by three core questions: (i) Can RGBC effectively distinguish between brain states (awake, slow oscillations, and microarousals) based on the complex, non-stationary nature of LFPs? (ii) How does RGBC's performance compare to that of state-of-the-art neural networks, particularly in terms of classification accuracy, computational efficiency, data requirements, and computational load?

To answer these questions, we implemented RGBC to decode brain states from invasive LFP recordings, extending its application beyond non-invasive signals. We benchmarked RGBC against CNNs and Euclidean-based classifiers using neural data recorded from multiple brain regions. Our results show that RGBC not only outperforms deep learning-based methods in classification accuracy but also offers remarkable improvements in computational efficiency, requiring far less training time, labeled samples, and computational resources.

## Materials and methods

### Overview of the framework

In this work, we present a simple and effective Riemannian geometry-based approach for brain state decoding, utilizing covariance matrices derived from multi-region intracortical electrophysiological recordings. Covariance matrices belong to the space of symmetric positive definite (SPD) matrices, which form a differentiable manifold with a curved, non-Euclidean structure (for a review, see (34)). Since conventional Euclidean metrics do not adequately capture the intrinsic geometry of this space, we introduce a Riemannian structure to facilitate precise distance computations, geodesic interpolation, and statistical analyses, which are essential for Minimum Distance to Mean (MDM)-based classification.

For the classification task, we used in vivo cortical local field potential (LFP) recordings from chronically implanted rats with a fixed number of electrodes ($\geq 3$) (35). Data were collected while the animals transitioned between different dynamic brain states (classes) in the induction and emergence from anesthesia. Each session was segmented into 3-second samples, and for each sample, we computed covariance matrices to capture inter-channel neural relationships, which served as features for classification. To classify brain states, we calculated the Minimum Distance to Mean (MDM), defined within a Riemannian framework, to assign each sample to the class whose mean covariance matrix was closest. Our proposed framework highlights the effectiveness of incorporating the geometric structure of the data for improved brain state decoding. For an overview of the dataset and the proposed framework, see **Fig.1**.

### Dataset

The dataset used in this study, first described in (35), consists of intracortical LFP signals recorded from Lister-Hooded rats (n=4) before, during, and after recovery from anesthesia (for a detailed description of the protocol, please refer to (35)). This dataset includes three main dynamic brain states: awake (AW), slow oscillations (SO), and microarousals (MA). These states were labeled by an expert experimentalist using the channel with the highest signal-to-noise ratio (SNR) as a reference. The temporal dynamics of these states are illustrated in **Fig.1**.

LFP signals were acquired using intracortical electrodes placed in various cortical regions, including prelimbic, parietal, visual, somatosensory, motor, and auditory areas. The dataset includes recordings from different combinations of cortical regions for each rat: Subject 1 had recordings from three regions (prelimbic, parietal, and visual), Subject 2 from five regions (prelimbic, parietal, visual, motor, and auditory), Subject 3 from four regions (prelimbic, parietal, visual, and somatosensory), and Subject 4 from five regions (prelimbic, parietal, visual, motor, and auditory). All sessions were videotaped, and LFPs were digitized at 10 kHz, and down sampled to 1 kHz for analysis, with recordings spanning two different days per rat.

*Processing and sample extraction*

For sample extraction, we divided the LFP recordings into 3-second windows, ensuring that each window contained data from a single brain state. These windows were treated as distinct sub-recordings, each representing a unique state/class. To ensure a balanced representation of brain states (classes) across all subjects and sessions, we counted the number of 3-second segments extracted for each state per subject and session. A minimum threshold of 200 segments per brain state was set, ensuring at least 600 segments (200 per state) were extracted from each session. For sessions with more than 200 segments per state, we randomly selected samples to maintain uniformity and prevent overrepresentation of any particular state.

*Manifold preparation and covariance estimation*

Before passing the samples through the MDM classifier, they were converted to a format suitable for geometric analysis. Each sample $v_i$, was converted into a covariance matrix $C_i \in R^{E \times E}$:

$$C_i = \text{Cov}_{\text{OAS}}(v_i), \qquad (1)$$

where E is the number of channels and $\text{Cov}_{\text{OAS}}$ is the covariance matrix for each sample using the Oracle Approximating Shrinkage (OAS) estimator (36). The OAS estimator was used due to its effectiveness with small sample sizes. This method not only reduces the bias inherent in small sample size covariance estimates but also adjusts the shrinkage intensity based on the variability observed within the sample. These enhancements are particularly beneficial in settings where the number of signal observations is limited or when the data contains high levels of noise, ensuring that the derived covariance matrices are both reliable and representative of the underlying neural processes. The collection of all covariance matrices forms a set $\mathcal{C} = \{C_i\}$, which was used as input for the MDM classifier in both the Riemannian and Euclidean frameworks.

*Euclidean and Riemannian geometry*

A spatial covariance matrix is inherently symmetric, and given sufficient data, it is also positive definite. These symmetric positive-definite (SPD) matrices form a differentiable manifold $P(E)$ of dimension $E *= E(E + 1)/2$. Particularly, when we provide an appropriate metric on it, the SPD space forms a Riemannian manifold enabling accurate distance and statistical metrics required for MDM classification tasks.

Neglecting the nonlinear geometry of the manifold $P(E)$ and treating it as a linear space allows us to apply Euclidean geometry methods directly (27,28). Under this assumption, a natural inner product in the space of symmetric matrices $S(E)$ is defined as:

$$\langle \cdot, \cdot \rangle : S(E) \times S(E) \to \mathbb{R},$$

$$(S_1, S_2) \mapsto \text{tr}(S_1, S_2),$$

with the associated Frobenius norm $||S||_F = \sqrt{tr(S^2)}$. In this Euclidean framework, the distance between two matrices $C_1, C_2 \in P(E)$ is given by:

$$d_{Euclid}(C_1, C_2) = ||C_1 - C_2||_F, \tag{2}$$

and the Euclidean mean of a set of covariance matrices $\{C_1, \ldots, C_{N_k}\}$ within a class $k$ is computed as:

$$G^k_{Euclid} = \frac{1}{N_k} \sum_{i=1}^{N_k} C_i.$$

This approach, however, ignores the $P(E)$ manifold inherent nonlinearities. A more suitable approach accounts for the Riemannian geometry of the previously defined manifold $P(E)$ by introducing a Riemannian metric. At each point $C \in P(E)$, the inner product in the associated tangent space $\mathcal{T}_C P(E)$ is defined as:

$$\langle \cdot, \cdot \rangle_C : \mathcal{T}_C P(E) \times \mathcal{T}_C P(E) \to \mathbb{R}$$
$$(S_1, S_2) \mapsto tr(C^{-1} S_1 C^{-1} S_2), \tag{3}$$

with the associated Riemannian norm $||S||_C = \sqrt{<S,S>_C}$.

In this Riemannian framework, the distance between two matrices $C_1, C_2 \in P(E)$ is given by the infimum of the length of smooth curves connecting them (37):

$$d_{Riem}(C_1, C_2) = inf\{\mathcal{L}(\gamma) \, | \, \gamma : [0,1] \to P(E), with \, \gamma(0) = C_1, \gamma(1) = C_2\},$$

where the length of a curve $\gamma$ is computed as:

$$\mathcal{L}(\gamma) = \int_0^1 ||\dot{\gamma}(t)||_{\gamma(t)} dt. \tag{4}$$

In a Riemannian manifold, a geodesic is defined as the unique curve of minimal length that connects $C_1$ and $C_2$. Particularly, in the manifold $P(E)$ equipped with the inner product, the geodesic between two points $C_1$ and $C_2$ and is given by (38):

$$\gamma(t) = C_1^{1/2} \left( C_1^{-1/2} C_2 C_1^{-1/2} \right)^t C_1^{1/2}, t \in [0,1].$$

This geodesic is derived using the matrix exponential map, which locally projects elements of the tangent space back onto the manifold. It smoothly interpolates between $C_1$ and $C_2$ while preserving the structure of the manifold. The Riemannian distance along this geodesic simplifies to:

$$d_{Riem}(C_1, C_2) = \left|\left| log \left( C_1^{-1/2} C_2 C_1^{-1/2} \right) \right|\right|_F = \left[ \sum_{i=1}^{E} log^2 \lambda_i \right]^{1/2}, \tag{5}$$

where $\{\lambda_i\}_{i=1}^{E}$ are the eigenvalues of $C_1^{-1/2} C_2 C_1^{-1/2}$ (or, equivalently $C_1^{-1} C_2$), ensuring symmetry.

In a Riemannian framework, given a set of covariance matrices $\{C_1, \ldots, C_{N_k}\}$ within a class $k$, the mean of the Riemannian class ($G^k_{Riem}$) is the Fréchet mean, defined as the point that minimizes the sum of squared Riemannian distances:

$$G^k_{Riem} = \arg\min_{C \in P(E)} \sum_{i=1}^{N_k} d^2_{Riem}(C, C_i).$$

Thus, this approach ensures that the geometric structure of the data is preserved, where distances and statistical metrics take into account the curvature of the space, which results in a more accurate representation of the data. Key properties such as congruence invariance (resistance to transformations like rotation), self-duality (similar treatment of matrices and their inverses), and determinant identity (accounting for the data's spread across dimensions) make this approach well-suited for analyzing covariance matrices in a meaningful way compared to traditional Euclidean methods (27,28).

*Minimum Distance to Mean Classifier*

The Minimum Distance to Mean (MDM) classifier computes the mean of covariance matrices for each class and assigns a test sample to the class whose mean is closest to it, based on a chosen distance. The chosen geometry approach determines the metric used in the MDM classifier. The Euclidean MDM classifier uses the Frobenius norm as the distance metric **(Eq.2)**, treating the space of SPD matrices as a linear Euclidean space, while the Riemannian MDM classifier uses the Riemannian distance **(Eq.5)**, which respects the nonlinear structure of the SPD manifold.

For a given test covariance matrix $C_{test}$, classification is performed by evaluating the geometric distance to each class mean $G^k_{metric}$ and assigning the label $\hat{y}$ of the closest class:

$$\hat{y} = \arg\min_k d_{metric}(C_{test}, G^k_{metric}), \tag{6}$$

where the pair $(G^k_{metric}, d_{metric})$ is determined based on the assumed geometry: $(G^k_{Euclid}, d_{Euclid})$ for the Euclidean geometry approach and $(G^k_{Riem}, d_{Riem})$ for the Riemannian geometry approach.

*Convolutional neural networks*

To benchmark the performance of the proposed MDM classifiers, we utilized a Convolutional Neuronal Network (CNN). CNNs are widely used for task classification and have been shown to achieve high accuracy in classifying brain states based on LFP signals, such as sleep states, anesthesia stages, and other similar tasks (24,25,39).

For the CNN, the original set of samples was used as input, unlike the Riemannian and Euclidean frameworks, which relied on covariance matrices. The only preprocessing step was standard scaling for normalization. The CNN architecture consisted of a convolutional stage followed by a dense stage. The convolutional stage included a 1D convolutional layer (Conv1D) with 32 filters, a 1D max pooling layer (MaxPooling1D), a second Conv1D layer with 64 filters, and a final 1D Max Pooling layer (MaxPooling1D) layer. Both Convolutional layers used same padding with *ReLU* activation, while the MaxPooling1D layers had a kernel size of 2 and *valid* padding. The convolutional stage ended with a flattening layer.

The dense stage consisted of a fully connected layer with 64 neurons and *ReLU* activation, followed by a final output layer. The number of neurons in the output layer matched the number of labels (three: AW, SO, and MA), with *softmax* activation to assign each input sample to the appropriate class. The model architecture remained unchanged regardless of the number of channels (regions) in the input data. We trained the model for 50 epochs.

*Benchmarking and validation*

The benchmarking process compared three approaches: the Riemannian MDM, the Euclidean MDM (used as a control), and CNNs (for comparison with conventional methods). The analysis was conducted in two rounds for each subject. In the first round, session 1 data was used for training and session 2 for testing, while in the second round, the setup was reversed. The primary performance metric was the macro-averaged F1 score. Final results were obtained by averaging the scores across all subjects and sessions.

To evaluate computational costs (computational efficiency), we used two hardware setups. For CNNs, models were trained and tested on both an NVIDIA A100 Graphics Processing Unit (GPU) and a Central Processing Unit (CPU). For the MDM approaches (both Riemannian and Euclidean), training and testing were performed exclusively on CPU resources. We recorded the time taken to train and test the models for each approach, enabling a comparison of the computational load across methods.

To evaluate the robustness of each approach, we tested model performance (accuracy) under limited data conditions by varying the size of the training set, using 60, 120, 300, and 600 trials. To ensure statistical reliability, we repeated the training and testing process 100 times to account for randomness introduced by factors such as sample selection and classifier initialization. Additionally, for evaluating time performance, we set the number of iterations to 10 for each test to reduce variability in time measurements.

## Results

In this study, we aimed to evaluate the efficacy of geometric data approaches in a classification task. We used a dataset from rats (n = 4), implanted with three to five intracortical electrodes, undergoing a controlled anesthesia recovery protocol, where the animals dynamically transition through three distinct brain states, i.e., classes that we aimed to classify. Our approach involves applying a Riemannian classifier to assess the extent to which data geometry improves classification performance. For benchmarking, we compared our approach to a vector-based MDM classifier using Euclidean distance, as well as a state-of-the-art CNNs.

*Riemannian geometric classifier outperforms state-of-art methods*

To evaluate our Riemannian geometric-based algorithm's performance, we tested different configurations of recording channels and subject inclusion criteria. Specifically, we examined three-channel (prelimbic, parietal, and visual), four-channel (prelimbic, parietal, visual, and sensory), and five-channel (prelimbic, parietal, visual, motor, and auditory) configurations. While the three-channel configuration was designed to maximize the number of included subjects while maintaining multiple recording regions, the four- and five-channel configurations prioritized increasing the number of regions, even if that required excluding some subjects. For each configuration, performance was assessed using the average F1 score across all runs. Each subject underwent two validation rounds, and results were aggregated over 100 iterations per case. This evaluation framework applies consistently to all subsequent reported results. We analyzed covariance matrices for each subject, where each matrix was constructed independently. When using multiple channels, a single covariance matrix was generated by concatenating data from all selected channels.

In **Table 1**, we summarized the accuracy results of our framework, demonstrating that Riemannian MDM outperforms both Euclidean MDM and CNN in terms of F1 score across most configurations tested. Specifically, with three channels, Riemannian MDM achieved the highest F1 score of 0.753±0.151, compared to 0.706±0.202 for CNN and 0.549±0.212 for Euclidean MDM. A similar trend was observed in the four-channel configuration. In the five-channel configuration, CNN slightly outperformed Riemannian MDM; however, this came with an increase in instability, as measured by the standard deviation (going from 0.097 for Riemannian MDM to 0.146 for CNN, resulting in a ~50% increment). Indeed, Riemannian MDM demonstrated greater stability compared to the other frameworks independently of the configuration used. On average, Riemannian MDM consistently produced both higher and more stable results (0.746±0.128) than CNN (0.698±0.214) and Euclidean MDM (0.556±0.224).

Our finding that the Riemannian geometry approach surpasses CNNs in classification tasks aligns with previous studies (40,41). However, other research highlights comparable decoding performance between Riemannian-based methods and state-of-the-art CNNs (42). Notably, most of these studies rely on noninvasive EEG datasets, whereas our evaluation focuses on intracortical LFP brain signals, an important factor to consider when comparing results due to the distinct nature of these signals (16).

Next, we evaluated whether Riemannian geometry could offer better computational efficiency while maintaining performance. Specifically, we i) tested how the performance of each method depends on the training size, a critical factor in BCI research where limited training data is often a challenge (42); and ii) assessed the training times required by each method using both GPU and CPU. To evaluate how accuracy depends on training size, we used data from the three-channel configuration, as discussed above, since this setup maximizes information while including all subjects. The training size ranged from 60 to 600 samples, with a step of 60 samples, while maintaining the same label distribution as in previous analyses (one-third per state; see Methods for details).

By evaluating performance based on training size, we demonstrated that the Riemannian MDM outperforms both the Euclidean MDM and the CNN (**Fig.2A**). Additionally, we showed that Riemannian MDM exhibits minimal dependence on training size, maintaining high F1 scores even with small amounts of data. In contrast, despite CNN achieving high accuracy, it required a sufficient amount of data to do so, performing worse when less data was available. Indeed, it is not new that CNN are sensitive to training data size (43), and techniques such as data augmentation are often employed to overcome this limitation (44). This fact underscores the potential of geometric machine learning, such as our proposed framework, which achieved high accuracy even with reduced training size, addressing a key challenge when working with smaller datasets (18).

As shown in **Fig.2B**, as the size of the training set increased, MDM methods demonstrated better scalability, maintaining lower training times compared to CNN. This difference became more pronounced as the training size became larger. When comparing the performance numerically (**Table 2**), for the smallest dataset size tested (training size = 10), we observed a ratio of 12.6x between the CNN trained on a GPU and the Riemannian MDM, which increased to 23.5x when compared to the CNN trained on a CPU. This ratio increases as the number of training samples increases, reaching a ratio of 61.5x for the CNN trained on a GPU, and 478.0x when CNN was trained on a CPU, for the largest size trained (training size = 10k). Note also that although the Euclidean MDM was slightly faster than the Riemannian MDM (0.015s vs. 0.227s for 10k samples, respectively), this difference was far outweighed by the Riemannian MDM's classification performance as shown in **Fig.2A**.

In summary, our results highlight the effectiveness of Riemannian MDM for efficient classification tasks, demonstrating both accuracy and speed with low computational demands. These characteristics are particularly valuable for large-scale neural decoding tasks, where computational efficiency is crucial (17,45).

*Riemannian Geometry methods' performance is state dependent across different brain regions*

Since Riemannian MDM outperformed other frameworks in both classification accuracy and computational efficiency, we next examined its sensitivity when using only pairs of brain regions rather than integrating data from multiple regions. To this end, with the region used for data labeling, we created pairs of brain regions. For visual inspection of our method's ability to discriminate among brain states (classes), we applied Uniform Manifold Approximation and Projection (UMAP) to the covariance matrices, comparing the ground truth labels with those obtained from our framework. For classifications accuracy, we calculated F1 scores as previously.

As illustrated in **Fig.3**, the UMAP projections of the ground truth data reveal that different pairs of brain regions vary in their ability to form well-separated clusters. For example, compare the cluster formed using the Prelimbic region to that formed using the Visual area (top-left vs. top-right in **Fig.3**). This difference is reflected in the classification accuracy, with the former outperforming the latter in distinguishing SO from the MA state. Notably, classification accuracy for the awake states was high regardless of the pairs of brain regions pair used. By examining the UMAP projections across different brain pairs, it becomes clear that the awake state forms a well-defined cluster, whereas the SO and MA tend to overlap. This highlights the relation between the geometric properties of the data and state dependence.

Overall, the Riemannian MDM captures smoother transitions between states while maintaining clear separability. By focusing on the underlying dynamics of neural activity, the Riemannian approach adapts to the state-dependent nature of the data, providing a more consistent representation of stable and continuous states. In contrast, the Euclidean MDM produces noisier embeddings with overlapping and less-defined clusters, highlighting its limitations in capturing neural state transitions (**Fig.3**).

## Discussion

In the current work, we investigate the efficacy of using geometry-based methods to classify cortical brain patterns. To do so, we explored the covariance structure of the data by employing a Riemannian approach. Unlike Euclidean space, where data lie in a flat vector space and computations rely on linear operations, the Riemannian approach treats covariance matrices as points on a curved manifold, preserving intrinsic geometric structure of the data, and ensuring that operations such as distance computation and classification respect the underlying topology (38). Based on this framework, we developed an MDM (Minimum Distance to Mean) classifier and demonstrated its efficacy in decoding brain states.

Riemannian geometry has been successfully applied to non-invasive brain recordings, such as EEG, fNIRS, or fMRI (27,28,31,32,46). In this work, we extend its application to intracortical LFP recordings which measure the brain activity in the extracellular space, and not at the scalp level as EEG (16). Our dataset comprises recordings from rats before and during the recovery from anesthesia (**Fig.1**), a protocol used to observe the emergence and transitions of distinct dynamic brain states (34,47–52). By recording from multiple brain regions, we computed the covariance matrix and used it as an input feature for classification. Our results demonstrate that our approach not only achieves high classification accuracy but also reduces computational time and complexity relative to a Euclidean MDM classifier and a convolutional neural network (**Fig.2**).

Given the multi-area recordings, we also investigated the model's ability to capture brain state dependencies using data from pairs of regions only. Visualization of the covariance matrices, embedded in a manifold via Uniform Manifold Approximation and Projection (UMAP), revealed that different brain states form distinct clusters (**Fig.3**). For example, the awake state tends to be well separated in this embedded space, while slow oscillations and microarousals tend to overlap (**Fig.3**). This observation might be attributed to the fact that both states, SO and MAs, share a slow frequency component that tends to synchronize the network, thus resulting in a similar position on the covariance matrices manifold (47). Indeed, the performance of the Riemannian classifier correlates with the degree of separability in the manifold, thereby revealing a state-dependent sensitivity. Moreover, the Riemannian classifier preserves the intrinsic geometry of the data, as evidenced by the lower performance of the Euclidean MDM classifier.

Overall, our findings reinforce the notion that geometry-based methods provide a robust framework for analyzing brain activity by leveraging the underlying data structure (42,53). This approach holds promise for various neurophysiological applications, including brain-state decoding and clinical assessments (55–57). In particular, it would be valuable to investigate how neural data is organized under pathological conditions, such as psychiatric disorders and brain lesions, where brain region interactions may be disrupted (58–62). Exploring the application of geometric methods in these contexts could offer novel insights into neural data organization in pathological states and reveal patterns associated with disease.

## Supporting information

The code used in the study will be made public with the manuscript publication.

## Acknowledgments

AML acknowledges the funding from RCC-Fellowship (rcc.harvard.edu/) from Harvard University. The authors acknowledge the European Union's Horizon 2020 Framework Programme for Research and Innovation under the Specific Grant Agreement No. 945539 (Human Brain Project SGA3) and by the ERC, NEMESIS, project number 101071900, PID2023-152918OB-I00 and AGAUR 2021-SGR-01165.

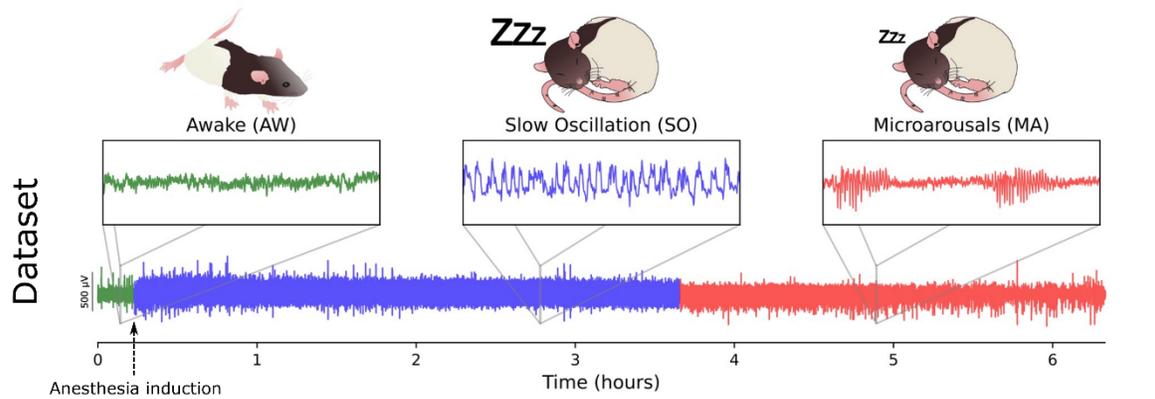

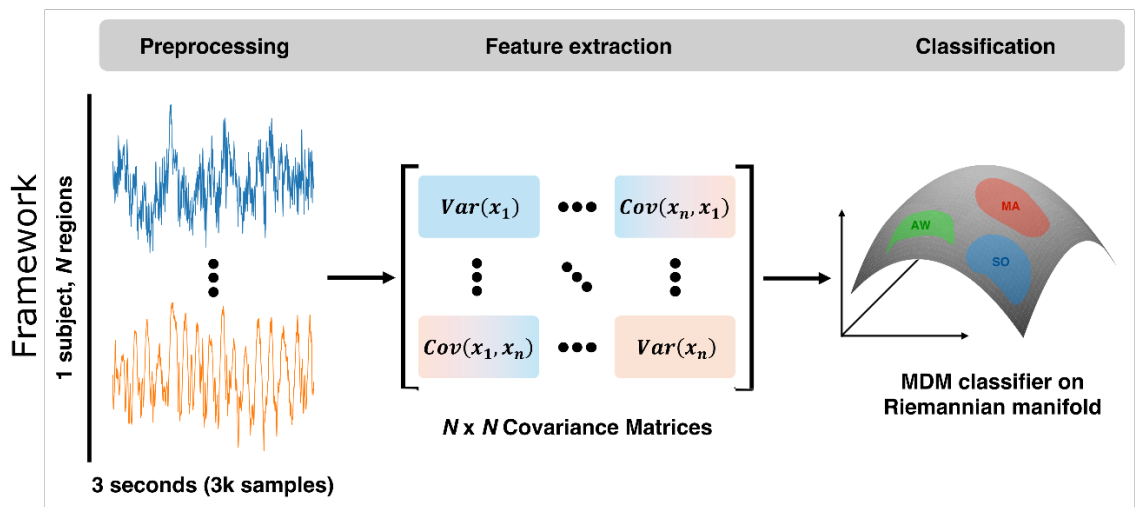

Figure 1 Schematics of the dataset and the proposed framework. Top: Local Field Potential (LFP) recordings were obtained before and during recovery from anesthesia, a process that revealed three distinct brain states: (1) AW, characterized by high-frequency and low-amplitude activity (green); (2) SO, associated with deep sleep or anesthesia, characterized by low-frequency and high-amplitude activity (blue); (3) MA, in which SO is interspersed with brief periods of AW (red). An example of the complete process is illustrated, showing the transitions and durations of each brain state. Bottom: Our framework segments the data into 3-second samples and computes covariance matrices to capture inter-channel neural relationships, which serve as features for classification. Classification is performed using the Minimum distance to mean (MDM) approach, where the Riemannian distance is employed to assign each sample to the class whose mean covariance matrix is closest.

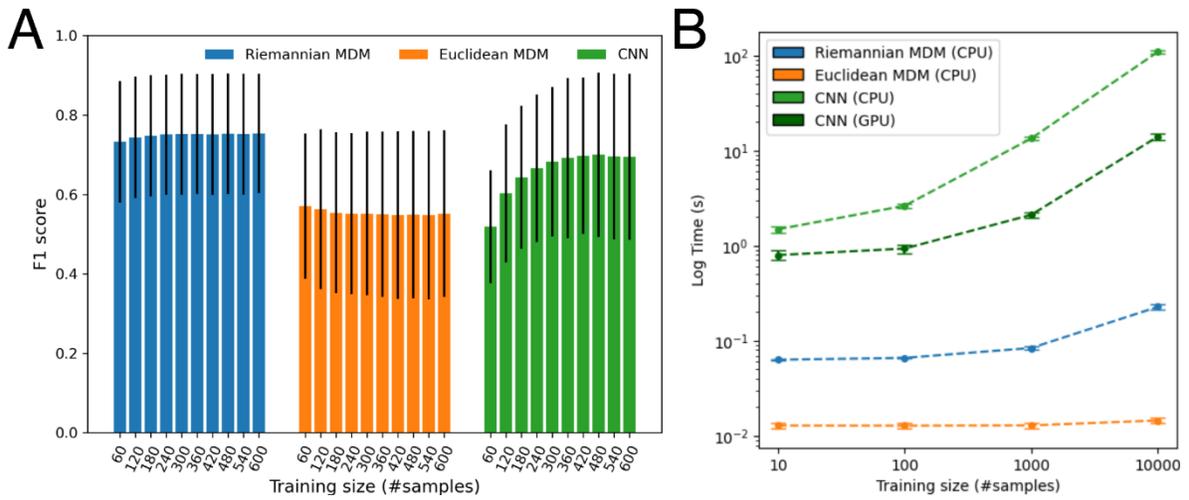

Figure 2 Evaluation of model performance across different frameworks. (A) Change in F1 score for all proposed frameworks as a function of the number of training samples. (B) Time performance changes for all proposed frameworks as a function of the number of training samples. Time axis in logarithmic scale. Mean ± 1 std.

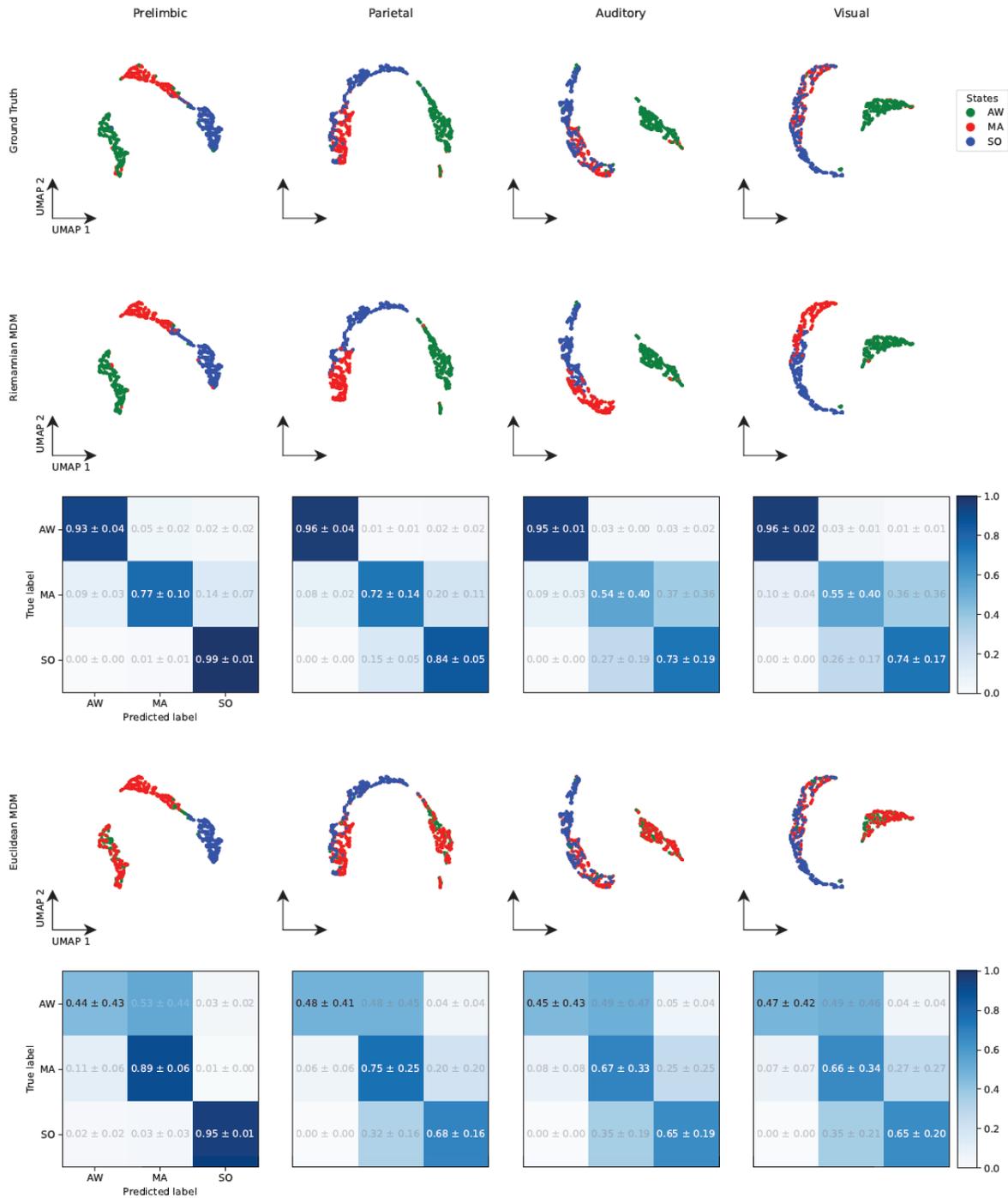

Figure 3 Riemannian geometry reveals brain-state separability and region-specific dynamics in LFP recordings. Each column corresponds to a specific brain region, from left to right: prelimbic, parietal, auditory, and visual cortex. Within each column, the top panel displays the Ground Truth (GT) UMAP projection of 2-second LFP windows, colored by annotated brain states (AW, SO, MA). The middle panel shows projections from Riemannian MDM decoding, while the bottom panel depicts projections from Euclidean MDM. Individual UMAP points represent LFP neural recording within 2-second windows, with adjacent confusion matrices quantifying classification consistency for each framework.

**Table 1.** F1 score comparing models' performance across different configurations and frameworks.

| #channels | #subjects | Framework | F1 score |
|---|---|---|---|
| 3 | 4 | Riemannian MDM | **0.753±0.151** |
|   |   | Euclidean MDM | 0.549±0.212 |
|   |   | CNN | 0.706±0.202 |
| 4 | 2 | Riemannian MDM | **0.679±0.094** |
|   |   | Euclidean MDM | 0.455±0.247 |
|   |   | CNN | 0.543±0.173 |
| 5 | 2 | Riemannian MDM | 0.806±0.097 |
|   |   | Euclidean MDM | 0.665±0.152 |
|   |   | CNN | **0.846±0.146** |
| Average | | Riemannian MDM | **0.746±0.128** |
|   |   | Euclidean MDM | 0.556±0.224 |
|   |   | CNN | 0.698±0.214 |

**Table 2.** Time performance by training size (number of samples) and framework.

| Training Size | Framework | Execution time (s) | Ratio with CNN CPU | Ratio with CNN GPU |
|---|---|---|---|---|
| 10 | Riemannian MDM (CPU) | 0.063±0.001 | 23.5x | 12.6x |
|   | Euclidean MDM (CPU) | **0.013±0.001** | **115.6x** | **61.9x** |
|   | CNN (CPU) | 1.488±0.120 | n/a | - |
|   | CNN (GPU) | 0.797±0.097 | 1.9x | n/a |
| 100 | Riemannian MDM (CPU) | 0.066±0.001 | 39.7x | 14.1x |
|   | Euclidean MDM (CPU) | **0.013±0.001** | **204.5x** | **72.6x** |
|   | CNN (CPU) | 2.627±0.153 | n/a | - |
|   | CNN (GPU) | 0.933±0.097 | 2.8x | n/a |
| 1000 | Riemannian MDM (CPU) | 0.084±0.002 | 159.2x | 25.1x |
|   | Euclidean MDM (CPU) | **0.013±0.001** | **1036.8x** | **163.6x** |
|   | CNN (CPU) | 13.397±0.558 | n/a | - |
|   | CNN (GPU) | 2.111±0.123 | 6.3x | n/a |
| 10000 | Riemannian MDM (CPU) | 0.227±0.014 | 478.0x | 61.5x |
|   | Euclidean MDM (CPU) | **0.015±0.001** | **7465.3x** | **961.2x** |
|   | CNN (CPU) | 108.739±3.309 | n/a | - |
|   | CNN (GPU) | 14.001±1.045 | 7.8x | n/a |